\documentclass[twocolumn,prl,floatfix,a4paper]{revtex4}

\usepackage[english]{babel}
\usepackage[utf8]{inputenc}
\usepackage[T1]{fontenc}

\usepackage{cancel,xcolor}
\usepackage{caption}
\usepackage{dsfont}
\usepackage{appendix}
\usepackage[normalem]{ulem}
\usepackage{enumitem}
\usepackage{blindtext}
\usepackage{comment}

\usepackage{braket}
\usepackage{amsmath}
\usepackage{amssymb}
\usepackage{amsfonts}
\usepackage{amsthm}
\usepackage{graphicx}
\usepackage[scale=0.8]{geometry}
\usepackage{float}

\definecolor{dark-green}{RGB}{0, 128, 0}

\newcommand{\hamo}{\mathcal{H}_0}
\newcommand{\ham}{\mathcal{H}_{\theta}}
\newcommand{\ii}{\mathrm{i}}
\newcommand{\ee}{\mathrm{e}}
\newcommand{\tr}{\mathrm{tr}}
\newcommand{\ch}{\mathrm{Ch}}
\newcommand{\dd}{\mathrm{d}}
\newcommand{\Eop}{\mathcal{E}}

\newcommand{\an}{\hat{a}}
\newcommand{\cre}{\hat{a}^\dagger}

\begin{document}

\title{Non-Hermitian spectral flows and Berry-Chern monopoles}
\author{Lucien Jezequel}
\affiliation{Ens de Lyon, CNRS, Laboratoire de physique, F-69342 Lyon, France}
\author{Pierre Delplace}
\affiliation{Ens de Lyon, CNRS, Laboratoire de physique, F-69342 Lyon, France}

\begin{abstract}
We propose a non-Hermitian generalization of the correspondence between the spectral flow and the topological charges of band crossing points (Berry-Chern monopoles). A class of non-Hermitian Hamiltonians that display a complex-valued spectral flow is built by deforming an Hermitian model while preserving its analytical index. We relate those spectral flows to a generalized Chern number that we show to be equal to that of the Hermitian case, provided a line gap exists. We demonstrate the homotopic invariance of both the non-Hermitian Chern number and the spectral flow index, making explicit their topological nature. In the absence of a line gap, our  system still displays a spectral flow whose topology can be captured by exploiting an emergent pseudo-Hermitian symmetry.
\end{abstract}
\maketitle

Non-Hermitian topology \cite{TopoBandNonHerm,NewNonHermiInvariants,TopoPhase} is an emergent topic stimulated by the rise of topological physics in various quantum and classical physical systems  \cite{lutopological2014,Nash14495,Delplace_2017, Shankar_2022, Zhang_2018, parker_2021}  that display -- naturally or on purpose --  non-Hermitian effects \cite{weimann_topologically_2017,GainLossExperiment,NonHermiPhysics,TopologicalLaser, NonlinearReview, Longhi2016}. 
Protected boundary states being certainly one of the most universal signatures in topological physics, the question of their existence in non-Hermitian systems has naturally led to numerous studies during the past few years.
This central issue of the bulk-boundary correspondence is however quite involved in non-Hermitian systems. The reason being that the spectrum of the periodic system from which a topological index is usually computed, and that of the system with open boundary conditions, can be totally distinct \cite{TonyLee16,xiong17}.
In particular, eigenstates of non-Hermitian open systems were found to be localized near the boundary, in contrast with Bloch waves of Hermitian models. This so-called non-Hermitian skin effect \cite{Yao2018} motivated the development of a non-Bloch bulk-boundary correspondence, where a generalized Brillouin zone is introduced to define the proper topological invariants in relation with the boundary states \cite{Yao2018, NonHemChernBands, Brzezicki19, Deng19, Borgnia20, guo2021analysis}. The calculation of this non-Hermitian Brillouin zone is however quite involved in itself, although recent analytical advances have been made recently to evaluate it in one dimension \cite{Yang2020}. Another approach consists in considering a complete biorthogonal basis, made of left and right eigenstates of the non-Hermitian Hamiltonian, in order to introduce a biorthogonal polarization that accounts for the appearance/disappearance of singular  edge modes in non-Hermitian systems with open boundaries  \cite{Kunst18,Edvardsson19,RestoBulkBoundary, Song_2019, Type34nonhermi}.

Instead of focusing on lattice problems and trying to adapt topological Bloch theory to the non-Hermitian realm, we address the issue of the non-Hermitian generalization of the correspondence between spectral flows and Berry-Chern monopoles, which are the topological charges associated to band crossing points \cite{VolovikBook, BerryChernMonopoles, Faure2019,Venaille2022}. 
In contrast with the bulk-boundary correspondence, the monopole-spectral flow correspondence does not involve open boundary conditions, but requires instead a variation in space of a physical quantity, such as a mass term or a vector potential. Moreover, when such a quantity varies linearly along a spatial direction -- say $x$ -- all the 'bulk' states are already localized around $x=0$ in the Hermitian case.

Berry-Chern monopoles are abundant in physics and their associated spectral flows have different physical interpretations depending on the system at hand. For instance, they arise in the low energy description of smooth interfaces between topologically distinct two-dimensional (2D) topological insulators \cite{Bellissard95, BerryChernMonopoles, touchais22} and metals \cite{upreti20}, they allow a suitable description of topological waves in various inhomogeneous continuous media \cite{Delplace_2017, perrot2019topological, Marciani_2020, venaille21, zhu2021topology,Langmuircyclotron, PerezPRL2022}, they account for the topological reorganisation of quantum levels in molecules \cite{iwai_topological_2014, iwai_2016,FZ2000,FZ2001}, and they also characterize the topology of 3D Weyl semi-metals \cite{Zyuzin_2012,Burkov_2015} and their generalizations \cite{Bradlyn2016, Ezawa2017}.

Despite its ubiquity, the question of the monopole-spectral flow correspondence has been overlooked in the context of non-Hermitian physics, leaving us with the following key questions: Does the spectral flow  survive non-Hermiticity? Is the correspondence between the spectral flow and the topological charge still valid, or does it break like the bulk-edge correspondence?
In this letter, we show how to construct a class of non-Hermitian models that preserve  the spectral flow. We then show how to extend its topological description, and build different mappings to relate it with the Berry-Chern monopoles of Hermitan models.

To do so, let us consider the generic two-fold band crossing Hamiltonian with a linear dispersion relation in the three directions $\lambda,x$ and $p$ 
\begin{align}
H_0[\lambda;x,p] =  \frac{1}{\sqrt{2}} \begin{pmatrix}
\lambda  & x- \ii p \\
 x+\ii p  &  -\lambda 
\end{pmatrix} \, .
\label{eq:H0}
\end{align}
The variables $x$ and $p$ must be understood as two canonical conjugate classical (commuting) observables, while $\lambda$ is a control parameter. The Hamiltonian \eqref{eq:H0} displays two bands of energy $E^{\pm} = \pm\sqrt{x^2+p^2+\lambda^2}$ which are separated by a gap except at the origin $\lambda=x=p=0$ where they touch. It is well known that such a degeneracy point constitutes a source of Berry curvature, whose flux through a surface $\mathcal{S}$ enclosing the origin in $(\lambda,x,p)$-space is a topological index called the first Chern number $\mathcal{C}_0$ \cite{Avron1989, VolovikBook, wan2011}, hence the name Berry-Chern monopole. This Chern number can be computed for each band as 
\begin{equation}
    \mathcal{C}^{\pm}_0 = \frac{1}{2\pi i} \int_{\mathcal{S}} \tr(P_0^{\pm} \dd P_0^{\pm} \wedge \dd P_0^{\pm})
    \label{eq:chern}
\end{equation}
where  $P_0^{\pm} = \ket{\psi^\pm}\!\! \bra{\psi^\pm}$ is the spectral projector on the positive/negative band, with $\ket{\psi^{\pm}}$ the two eigenstates of $H_0$ and $\dd P_0^{\pm}$ is the 1-form $\partial_\lambda P_0^{\pm} \dd\lambda+\partial_x P_0^{\pm} \dd x+\partial_p P_0^{\pm} \dd p$. In that case the Chern numbers of the bands are equal to $\mathcal{C}^\pm =\pm 1 $.
This non-zero value characterizes the impossibility to define a smooth gauge for the eigenstates of $H_0$ over the parameter space. In more formal words, it is a topological property of the $U(1)$-fiber bundle over the base space $S^2$. This bundle terminology simply means that at each point $(\lambda,x,p)$ of the base space, the eigenstates $\ket{\psi^\pm}$ are defined up to a phase, owing to their normalization which is preserved due to Hermiticity of $H_0$. Extending this topological property to non-Hermitian Hamiltonians is therefore not obvious, precisely because this $U(1)$-fiber bundle structure is, in general, lost. 

To circumvent this difficulty, we interpret $H_0[\lambda;x,p] $ as the \textit{symbol Hamiltonian}, or semiclassical limit, of the \textit{operator Hamiltonian} \cite{Faure2019, BerryChernMonopoles, Venaille2022}, 
\begin{align}
\hamo[\lambda] = \begin{pmatrix}
\lambda  & \cre \\
 \an  &  -\lambda 
\end{pmatrix}
=\an \sigma_- + \cre \sigma_+ + \lambda \sigma_z
\end{align}
where $\sigma_\pm = \sigma_x \pm \ii \sigma_y$ with $\sigma_i$ $(i=x,y,z)$ the Pauli matrices, and $\an$ and $\cre$ are the bosonic annihilation and creation operators that satisfy $[\an,\cre]=1$. Those operators are related to the (non-commuting) position and momentum operators that satisfy $[\hat{x},\hat{p}]=\ii$ ($\hbar=1$), as $\hat{a} = (\hat{x}+i\hat{p})/\sqrt{2}$, $\hat{a}^\dagger = (\hat{x}-i\hat{p})/\sqrt{2}$. Up to the rescaling $\lambda/\sqrt{2}\rightarrow\lambda$, the symbol Hamiltonian \eqref{eq:H0} is recovered  by simply replacing the operators $\hat{x}$ and $\hat{p}$ by their classical commuting counterpart $x$ and $p$. Note that the same model is recovered for a Weyl fermion in a magnetic field where the  magnetic momenta in the orthogonal plane become canonical conjugated observables (such as $\hat{x}$ and $\hat{p}$), while the longitudinal momentum plays the role of $\lambda$ \cite{BerryChernMonopoles}. 

\begin{figure*}[t!]
\centering
\includegraphics[width=1\textwidth]{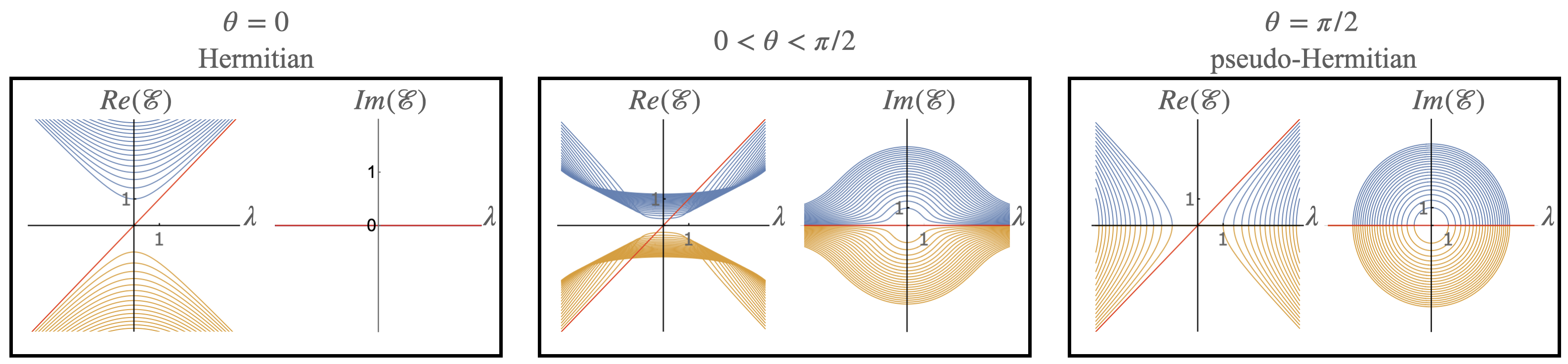}
\caption{\label{fig:spectra_nonHermitian} Real and imaginary part of the eigenvalue spectrum $\Eop^{\pm}_{\theta,n}$ of $\ham$ as a function of $\lambda$, up to $n=20$, for $\theta=0,5\pi/12,\pi/2$. The spectral flow (in red) is unaffected by the non-Hermitian $\theta$-deformation. }
\end{figure*}

As shown in figure \ref{fig:spectra_nonHermitian} (up-left), the spectrum of $\hamo$ is made of two branches $\pm$ given by the set
\begin{align}
    \label{eq:spectrum_H0}
    \Eop_{0,n}^{\pm} &= \pm \sqrt{\lambda^2 + (n+1)} \quad \quad n\in \mathbb{N}  \\ 
    \Eop_{-1} &= \lambda   \ .
\end{align}
The additional mode of energy $\Eop_{-1}=\lambda$ constitutes the \textit{spectral flow}, as it transits from the negative to the positive branch when $\lambda$ varies from $-\infty$  to $+\infty$. In other words, the positive branch gains $\mathcal{N}^+=1$ mode while the negative branch gains  $\mathcal{N}^-=-1$ mode. It is a remarkable result that those two numbers are precisely given by the Chern numbers \eqref{eq:chern} of the symbol Hamiltonian \eqref{eq:H0} as $\mathcal{N}^\pm=\mathcal{C}^\pm$. This relation is precisely the monopole-spectral flow correspondence.
Importantly, for that model, the spectral flow is directly accounted by the analytical index
\begin{align}
 \text{ind} \mathcal{D} \equiv \text{dim Ker}\mathcal{D}- \text{dim Ker}\mathcal{D}^\dagger  = \mathcal{N}^+
 \label{eq:index}
\end{align}
with $\mathcal{D}=\an \sigma_-$ \cite{BerryChernMonopoles}. The relation between this analytical index and the Chern index of the symbol Hamiltonian is known as the index theorem \cite{atiyah1968index, nakahara2018geometry}.

Remarkably, the analytical index \eqref{eq:index} is defined irrespective of the diagonal part of $\hamo$ and remains also unchanged  when multiplying the off-diagonal elements $\mathcal{D}$ and $\mathcal{D}^\dagger$ by an arbitrary complex number. Those two crucial points allow us to deform $\hamo$ in two different  non-Hermitian ways (i.e. diagonal and non-diagonal) that leave the analytical index invariant.
We are thus led to introduce the non-Hermitian operator Hamiltonian
\begin{align}
\mathcal{H}[\lambda] = \begin{pmatrix}
\lambda \ee^{\ii \varphi} & z_\alpha \cre \\
z_\beta \an  &  -\lambda \ee^{\ii \varphi}
\end{pmatrix}
\label{eq:H_op}
\end{align}
with $z_{\alpha/\beta}\in \mathbb{C}$ and $\varphi$ a phase,  that has the same analytical index as $\hamo$. All the parameters introduced in \eqref{eq:H_op} break Hermiticity, but they actually do not play the same role. Indeed, one can reduce the analysis of this Hamiltonian to  that of the much simpler one
\begin{align}
\ham[\lambda] = \begin{pmatrix}
\lambda & \ee^{\ii \theta} \cre \\
\ee^{\ii \theta} \an & -\lambda
\end{pmatrix}    
\end{align}
through the transformation
\begin{align}
    A^{-1} \mathcal{H}[\sqrt{r_\alpha r_\beta}\lambda] A = \sqrt{r_\alpha r_\beta}\ee^{\ii \varphi} \ham[\lambda] 
    \label{eq:transformA}
\end{align}
where we have introduced $\theta\equiv (\varphi_\alpha+\varphi_\beta)/2-\varphi$, with $z_{\alpha/\beta}=r_{\alpha/\beta}  \ee^{\ii\varphi_{\alpha/\beta}}$ and 
\begin{align}
    A = \begin{pmatrix}
   z_\alpha^{1/2} & 0 \\
    0 & z_\beta^{1/2} 
    \end{pmatrix} \ . 
\end{align}
Note that $\ham$ corresponds to the Hermitian Hamiltonian $\hamo$ introduced above when $\theta=0$. The non-unitary transformation \eqref{eq:transformA} means that if $\ket{\psi(\lambda)}$ is an eigenstate of $\ham[\lambda]$ with the eigenvalue $\Eop_\theta(\lambda)$, then $A\ket{\psi(\lambda/\sqrt{r_\alpha r_\beta})}$ is an eigenstate of $\mathcal{H}[\lambda]$ with the eigenvalue $\sqrt{r_\alpha r_\beta} \ee^{\ii \varphi} \Eop_\theta(\lambda/\sqrt{r_\alpha r_\beta})$.
So, up to a re-scaling by a factor $\sqrt{r_\alpha r_\beta}$ and to  a global rotation of angle $\varphi$ of the spectrum in the complex plane, the study of $\mathcal{H}$  reduces to that of $\ham$, whose spectrum reads (see Supplemental Material (SM))
\begin{align}
    \Eop_{\theta,n}^{\pm} &= \pm \sqrt{\lambda^2 + \ee^{2\ii \theta} (n+1)} \quad \quad n\in \mathbb{N}  \label{eq:spectrum_Eop_theta} \\ 
    \Eop_{-1} &= \lambda \ .
    \label{eq:spectrum_Eop_sf}
\end{align}
The asymptotic behaviour $\Eop^{\pm}_{\theta,n} \rightarrow \pm$sgn$(\lambda)$ for each $n$ and every $\theta$ when $|\lambda|\rightarrow \infty$ implies that $\Eop_{-1} \rightarrow \Eop^{\pm}_{\theta,n}$ when $\lambda\rightarrow \pm \infty$. This means that the operator Hamiltonian $\ham$ (and therefore $\mathcal{H}$) displays a spectral flow with respect to the parameter $\lambda$. 
Examples are  shown in figure \ref{fig:spectra_nonHermitian} for different values of $\theta$, and additional spectra of $\mathcal{H}$ are shown in SM.

\begin{figure}[t]
\centering
\includegraphics[width=0.49\textwidth]{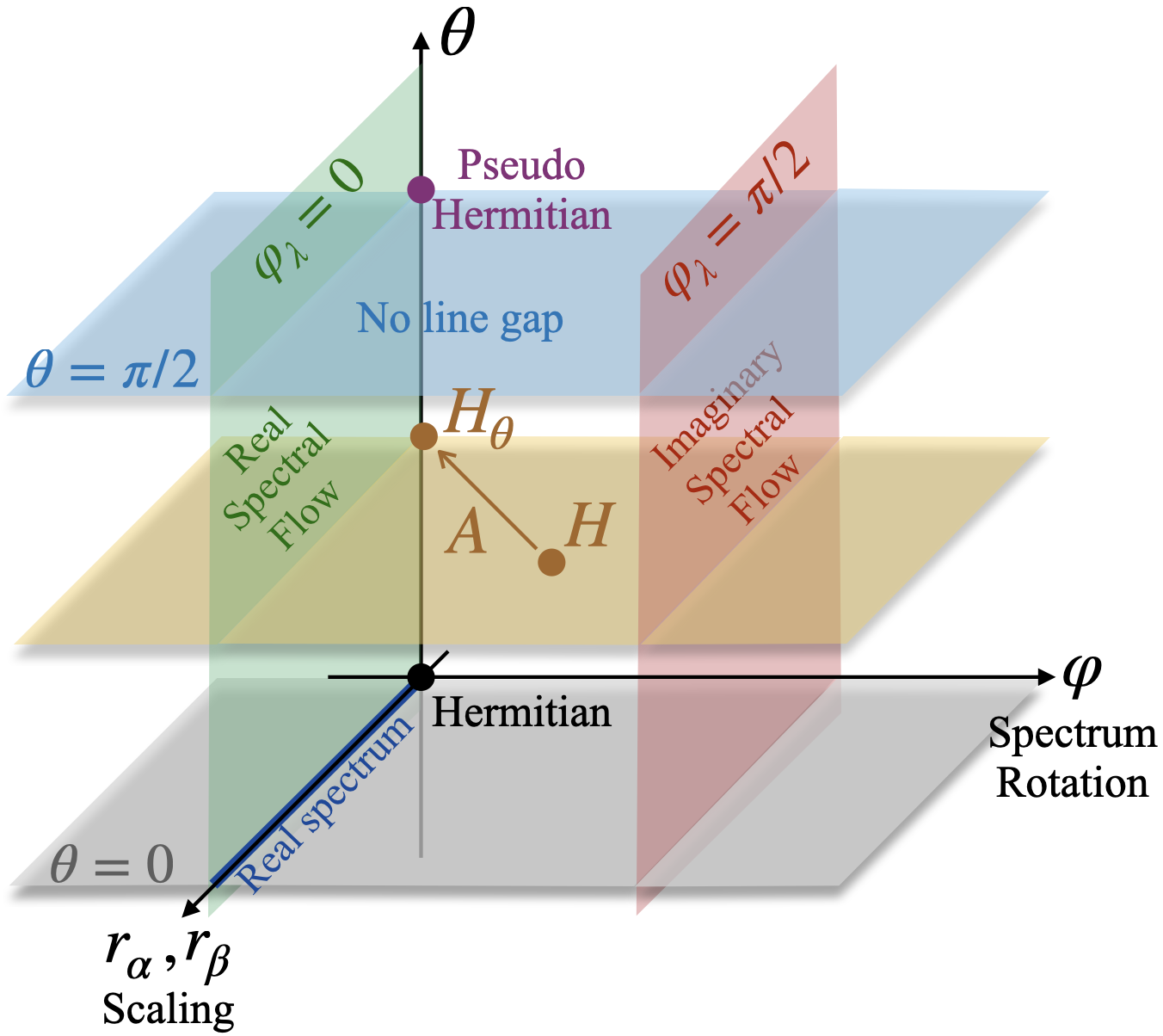}
\caption{\label{fig:regimes} Diagram of the different parameters that break Hermiticity. The origin, where the model becomes Hermitian, is taken as $(r_\alpha,r_\beta,\varphi,\theta)$ $=$ $(1,1,0,0)$. The transformation $A$ ''projects'' the non-Hermitian Hamiltonian $H$ into $H_\theta$. Under this transformation, the model becomes Hermitian  when $\theta=0$, and pseudo-Hermitian when $\theta=\pi/2$. Whenever $\theta=\pi/2$, there is no line gap in the spectrum of the symbol Hamiltonian but the operator Hamiltonian still displays a spectral flow. The spectral flow is in general complex-valued, except for $\varphi=0$ and $\varphi=\pi/2$ where it becomes purely real or purely imaginary respectively.    }
\end{figure}

Figure \ref{fig:regimes} summarizes the role of the different parameters that break Hermiticity in the full model. 
A first important remark is that for $\varphi=0$ modulo $\pi$ (and thus in particular for $\ham$), the spectral flow eigenvalue $\mathcal{E}_{-1}=\lambda$ remains purely real, unlike the rest of the spectrum which is in general complex-valued.  We  furthermore find  that the corresponding eigenmode remains localized around $x=0$ (see SM). Therefore, the spectral flow mode of $\hamo$ remains invariant under the action of all the  non-Hermitian terms introduced in \eqref{eq:H_op} but the parameter $\varphi$ which rotates its energy.
Interpreting the spectral flow as a chiral mode at a smooth interface between two Chern insulators, this result contrasts that of \cite{TopoBandNonHerm} where the interface state's localization length is found to be affected by the non-Hermiticity at a sharp step-like domain wall.

If now, in addition to $\varphi=0$, we also assume $\theta=0$, then the full spectrum of $\mathcal{H}$ becomes  real-valued, irrespective of the non-Hermitian asymmetry  $r_\alpha/r_\beta \neq 1$. Real spectra are an unusual property of non-Hermitian matrices, and their existence in physical systems stimulated recent works \cite{RussellYang22,Realspecwithnosym}. Here, this property is explained by the transformation \eqref{eq:transformA} that maps, in that case,  $\mathcal{H}$ onto the Hermitian operator $\hamo$.    

Finally, $\ham$ displays a striking spectrum for $\theta=\pi/2$. There, the spectral flow does not bridge two branches separated by a spectral gap in energy, but relates instead two branches separated by a range in $\lambda$ where the energy of the modes become purely imaginary and self-conjugated.
As we detail below, this case, that requires a specific treatment, can be understood as a pseudo-Hermitian symmetry breaking phase.

For now, let us focus on the interpolation between the Hermitian case $\theta=0$ and the pseudo-Hermitian one $\theta =\pi/2$. As $\theta$  increases, the real part of the spectrum of  $\mathcal{H}_\theta$ ''unfolds'' around the spectral flow level $\mathcal{E}_{-1}$, while  the imaginary part grows from zero but remains finite, leaving the spectral flow intact (figure \ref{fig:spectra_nonHermitian}). This robustness can be interpreted topologically  from the  non-Hermitian symbol of $\ham$ that reads
\begin{align}
H_{\theta}[\lambda;x,p] =  \frac{1}{\sqrt{2}} \begin{pmatrix}
\lambda  & \ee^{\ii \theta}(x- \ii p) \\
 \ee^{\ii \theta}(x+\ii p)  &  - \lambda 
\end{pmatrix}  \ .
\label{eq:H}
\end{align}
Its spectrum $E^\pm =\pm  \sqrt{\lambda^2+\ee^{2 \ii \theta}(x^2+p^2)}$ preserves the degeneracy point at $\lambda=x=p=0$.
Owing to the lost of Hermiticity of $H_\theta$, one needs to find a non-Hermitian generalisation of the Chern number to capture its topological properties \cite{TopoBandNonHerm, WeylChern, NonHemChernBands, Type34nonhermi}.
Here we propose the quantity
\begin{align}
\mathcal{C}^{\pm}_{\theta}  \equiv \frac{1}{2\pi i} \int_{(\lambda,x,p) \in S^2} \tr(P^{\pm}_{\theta} \dd P^{\pm}_{\theta} \wedge \dd P^{\pm}_{\theta}) \label{eq:ChernnonHerm}
\end{align} 
where the $P^{\pm}_{\theta}\equiv\ket{\psi^{\pm}}\!\!\bra{\Tilde{\psi}^{\pm}}$ are the non-orthogonal spectral projectors of the complex energy bands $\pm$, with $\ket{\psi^{\pm}}$ and $\ket{\Tilde{\psi}^{\pm}}$  the right and left eigenstates of $H_{\theta}$ respectively. 
These projectors are well-defined as long as the two spectral bands $E^\pm$ are separated by a line in the complex plane (the so-called line-gap \cite{NonHermiPhysics}).

We then use the invariance by homotopy of $\mathcal{C}^{\pm}_{\theta}$ with respect to $\theta$ to show the equality between this generalized Chern number and the usual one defined in the Hermitian case, i.e. $\mathcal{C}^\pm_\theta = \mathcal{C}^\pm_0$,  as long as the line gap is preserved (see SM). As the spectral flow index \eqref{eq:index} is also invariant in the non-Hermitian regime (see SM), this extends the monopole-spectral flow correspondence to the non-Hermitian systems given by the operator Hamiltonian $\ham$ and its symbol $H_\theta$. This correspondence finally generalizes to the full non-Hermitian system defined by the operator Hamiltonian $\mathcal{H} $ given in \eqref{eq:H_op} and its symbol $H$, since the transformation \eqref{eq:transformA} preserves the Chern number (see SM). 

The existence of a line gap is required for the projectors $P^\pm_\theta$ to be well-defined. One can check that such a line gap indeed exists along the imaginary axis for $H_\theta$  (and is tilted by and angle $\varphi$ in the full problem) except for $\theta=\pi/2$, as sketched in figure \ref{fig:regimes}. In that case, the generalized Chern number \eqref{eq:ChernnonHerm} is ill-defined, and we thus need to come with another strategy to capture the topology of $H_{\pi/2}$.

For that purpose, let us notice that $H_{\pi/2}$ (as well as $\mathcal{H}_{\pi/2}$) owns the pseudo-Hermitian symmetry  $\sigma_z H_{\pi/2}[\lambda;x,p] \sigma_z=H^\dagger_{\pi/2}[\lambda;x,p]$. This symmetry implies that the eigen-energies are either real or appear as complex conjugate pairs. When a parameter is varied (e.g. $\lambda$ here), eigen-energies can switch from the first case to the other in what is called a spontaneous pseudo-Hermitian symmetry breaking \cite{NonHermiPhysics, Type34nonhermi, SymProtectedDelplace}, as observed here. This pseudo-Hermiticity allows us to map $H_{\pi/2}$ onto an Hermitian Hamiltonian for which the topology is well-defined (see details in the SM).

The idea behind this mapping is to notice that the iso-energy surfaces in parameters space for $\theta=\pi/2$ describe hyperboloids $E^2=\lambda^2-x^2-p^2$, in contrast with the usual Hermitian case $\theta=0$ where they describe spheres $E^2=\lambda^2+x^2+p^2$. Actually, the pseudo-Hermitian case also yields a sphere, but a different one, since we have $\lambda^2=E^2+x^2+p^2$. This hint suggests that one should change our point of view and consider our model as an eigenvalue problem in $\lambda$ in parameter space $(E,x,p)$. Thinking of $\lambda$ as a momentum in a $y$ direction, and of the energy $E$ as the quantum number associated to $\partial_t$, such a  transformation could be formally thought as a space-time $y \leftrightarrow t$ change of axes.

To do so, it is convenient to first perform a unitary transformation $U H_{\pi/2} U^\dagger = \tilde{H}_{\pi/2}$, that preserves the pseudo-Hermiticity, which, for the Hamiltonian operator yields 
\begin{align}
    \Tilde{\mathcal{H}}_{\pi/2}[\lambda]\equiv \begin{pmatrix}
\lambda & \cre\\ -\an&-\lambda\\
\end{pmatrix}
\end{align}
with $U=\ee^{\ii \frac{\pi}{4} \sigma_z}$. The eigenvalue problem in $\Eop$ for $\Tilde{\mathcal{H}}_{\pi/2}$ is actually equivalent to the following eigenvalue problem in $\lambda$ 
\begin{equation}
    \begin{pmatrix}\Eop&\cre\\ \an&-\Eop
    \end{pmatrix} \ket{\psi'}=\lambda \ket{\psi'} 
    \label{eq:lambda-E}
\end{equation}
with $\ket{\psi'} = -\sigma_z \ket{\psi}$. The equation \eqref{eq:lambda-E} is nothing but $\hamo[\Eop]\ket{\psi'}=\lambda \ket{\psi'}$, meaning that the eigenvalue problem in $\Eop$ parametrized by $\lambda$ at $\theta=\pi/2$, is the same as the Hermitian eigenvalue problem in $\lambda$, parametrized by $\Eop$. Thus, interchanging $\theta=\pi/2 \leftrightarrow \theta=0$ amounts to swapping the roles of $\Eop$ and $\lambda$ as parameters and eigenvalues, as it actually appears in figure \ref{fig:spectra_nonHermitian} when both $\Eop$ and $\lambda$ are real. 

The realness of $\lambda$, being the eigenvalue of $\hamo[\Eop]$ in \eqref{eq:lambda-E}, is ensured by imposing the realness of $\Eop$, since $\hamo[\Eop]$ is Hermitian in that case. 
By doing so, we only keep the eigenmodes of $\Tilde{\mathcal{H}}$ with a real energy and dismiss those in the spontaneously broken phase. As a consequence, the pseudo-Hermitian symmetry broken phase appearing in the Re$\, \Eop$ spectrum parametrized by $\lambda$, is now interpreted as a disappearance of eigenvalue, namely a gap in the spectrum in $\lambda$ parametrized by  Re$\, \Eop$.
Since the mapping \eqref{eq:lambda-E} is also valid for the symbol Hamiltonian $H_\theta$, we conclude that the spectral flow of $\mathcal{H}_{\pi/2}[\lambda]$ is captured by the Chern numbers of $H_{0}[\text{Re}\,E,x,p]$ whose value is $\pm1$. Indeed, this Hamiltonian is formally equivalent to the previously discussed symbol Hamiltonian $H_{0}[\lambda,x,p]$ after the substitution $\text{Re}\,E \leftrightarrow \lambda$, for which the Chern numbers take the values $\mathcal{C}_0=\pm 1$. In other words, the spectral flow of $\mathcal{H}_{\pi/2}[\lambda]$ is thus determined by the Chern numbers of the Hermitian monopole.

\textit{Summary}--
This work provides a non-Hermitian generalization of the correspondence between a spectral flow of an operator and the Chern numbers of the eigenstates bundle associated to the degeneracy point of the (local) symbol Hamiltonian. Our analysis has direct applications in the investigation of topological properties in various systems such as in photonics, fluids and plasmas where non-Hermitian effects are abundant.
Finally, the two-band model we have considered, seen as a spin $1/2$ model, can be generalized to higher spins and non-linear dispersion relations in future works, as it is known that  Hermitian spectral flows appear in such models \cite{Bradlyn2016, Delplace_2017, Ezawa2017, Marciani_2020, BerryChernMonopoles}.

\textit{Acknowledgment}-- L. J. was funded by a PhD grant allocation Contrat Doctoral Sp\'ecifique Normalien.

%\bibliographystyle{unsrt}
%\bibliography{bibliography}

%%%%%%%%%%%%%%%%%%%%%%%%%%%%%%%%%%%%%%%%%%%%%%%%%%%%%%%%%%%%
\section{Eigenmodes of $\ham$ and $\mathcal{H}$}

In this section, we determine the non-Hermitien eigenenergies and eigenmodes  of the Hamiltonian operator
\begin{align}
\ham[\lambda] = \begin{pmatrix}
\lambda & \ee^{\ii\theta} \cre \\
\ee^{\ii\theta} \an  &  -\lambda
\end{pmatrix}\ .
\label{eq:H_op_appendix}
\end{align}
To do so, let us introduce the  basis of \textit{number states} (or Fock states) $\ket{n}$ which consists of the excitation modes associated to the creation/annihilation operators satisfying
\begin{align}
    \cre\ket{n}&=\sqrt{n+1}\ket{n+1}\\
    \an\ket{n}&=\sqrt{n}\ket{n-1}
\end{align}
for $n\in \mathbb{N}$. Looking for eigenstates $\ket{\psi_n}$ of the form 
$( c_1 \ket{n+1}, c_2 \ket{n})^t$
where $c_1$ are $c_2$ are numbers depending on $n$, leads to
\begin{align}
\label{eq:c2c1}
c_2\ee^{\ii\theta} \cre \ket{n} &= c_1(\Eop - \lambda) \ket{n+1}\\
c_1\ee^{\ii\theta} \an \ket{n+1} &= c_2(\Eop+\lambda) \ket{n} \, .
\label{eq:c1c2}
\end{align}
Multiplying \eqref{eq:c1c2} by $ \ee^{\ii\theta}\cre$ and substituting \eqref{eq:c2c1}, one gets 
\begin{align}
 \ee^{\ii 2 \theta} \cre \an \ket{n+1} = (\Eop +\lambda)(\Eop - \lambda)    \ket{n+1} .
\end{align}
Then, using the relation $\cre\an \ket{n}=n\ket{n}$, one gets the  spectrum $\Eop_{n}^{(\pm)} = \pm \sqrt{\lambda^2+ \ee^{2\ii\theta} (n+1)}$ of the main text.

Next, the real space representation of the modes $\psi_n^{\pm}(x)=\braket{x|\psi_n^{\pm}}$ is obtained by using the usual expressions
\begin{align}
\cre &= \frac{1}{\sqrt{2}}(x-\partial_x)\\
\an &= \frac{1}{\sqrt{2}}(x+\partial_x)
\end{align}
that yield  $\braket{x|n}= C_n e^{-x^2/2}H_n(x)$ where the $H_n$'s are the Hermite polynomials and $C_n=\frac{1}{\sqrt{2^n n! \sqrt{\pi}}}$.
Finally, using e.g Eq.\eqref{eq:c1c2}, one infers a relation between $c_1$ and $c_2$, and find
\begin{equation}
\begin{aligned}
\psi_{n}^\pm(x) = C_{n+1}\ee^{-x^2/2}\begin{pmatrix}(\lambda  + \Eop_{n}^{(\pm)})\,  H_{n+1}(x)\\
\sqrt{2} \ee^{\ii\theta} (n+1)\, H_n(x)  \end{pmatrix}
\end{aligned}\ .
\end{equation} 
Those modes are localized around $x=0$, as in the Hermitian case, but are altered by the non-Hermitian term $ \ee^{\ii\theta}$. 
In particular the two modes $\psi_{n}^\pm$ stop being orthogonal to each other.

One can then check that the family $(\psi_{n}^\pm)_{n,\pm}$ almost spans the entire Hilbert space except for a last vector $ (
\ket{0},0)^t$,  which is also an eigenvector of $\ham[\lambda]$ and complete the basis. We denote this last mode by $\ket{\psi_{-1}}$ with a slight abuse of notation. This is the mode of energy $\Eop_{-1} = \lambda$ that contributes to the spectral flow. Remarkably, its spatial profile, 
\begin{equation}
\begin{aligned}
\psi_{-1}(x) = 
\begin{pmatrix}
\ee^{-x^2/2}\\ 0 
\end{pmatrix}
\end{aligned}
\end{equation}
that is also localised around $x=0$, is the only one that remains unaffected by the non-hermitian term $\ee^{\ii\theta}$ in sharp contrast with the other $\psi_n^{\pm}(x)$ modes.

Finally, the more general non-Hermitian Hamiltonian
\begin{align}
\mathcal{H}[\lambda] = \begin{pmatrix}
\lambda \ee^{\ii \varphi} & z_\alpha \cre \\
z_\beta \an  &  -\lambda \ee^{\ii \varphi}
\end{pmatrix}
\label{eq:H_opappendix}
\end{align}
can be expressed as
\begin{align}
     \mathcal{H}[\sqrt{r_\alpha r_\beta}\lambda]  = \sqrt{r_\alpha r_\beta}\ee^{\ii \varphi} A\ham[\lambda] A^{-1}
    \label{eq:transformAappendix}
\end{align}
for $\theta\equiv (\varphi_\alpha+\varphi_\beta)/2-\varphi$, with $z_{\alpha/\beta}=r_{\alpha/\beta}  \ee^{\ii\varphi_{\alpha/\beta}}$ and 
\begin{align}
    A = \begin{pmatrix}
   z_\alpha^{1/2} & 0 \\
    0 & z_\beta^{1/2} 
    \end{pmatrix} \ . 
\end{align}
therefore, after the substitution $\lambda \xrightarrow{}\sqrt{r_\alpha r_\beta}\lambda \equiv \tilde{\lambda}$, the energies of this Hamiltonian are rescaled as
\begin{align}
    \label{eq:spectrum_H0appendix}
    \Eop_{n}^{\pm} &= \pm \sqrt{r_\alpha r_\beta}\ee^{\ii \varphi}\sqrt{\tilde{\lambda}^2 + (n+1)} \quad \quad n\in \mathbb{N}  \\ 
    \Eop_{-1} &= \sqrt{r_\alpha r_\beta}\ee^{\ii \varphi}\tilde{\lambda}   \ .
\end{align}
and the eigenmodes are modified as
\begin{equation}
\begin{aligned}
A\psi_{n}^\pm(x) = C_{n+1}\ee^{-x^2/2}\begin{pmatrix}z_\alpha^{1/2}(\tilde{\lambda}  + \Eop_{n}^{(\pm)})\,  H_{n+1}(x)\\
z_\beta^{1/2}\sqrt{2} \ee^{\ii\theta} (n+1)\, H_n(x)  \end{pmatrix}
\end{aligned}
\end{equation} 
except for the spectral flow mode that becomes
\begin{equation}
\begin{aligned}
A\psi_{-1}(x) = 
z_{\alpha}^{1/2}\begin{pmatrix}
\ee^{-x^2/2}\\ 0 
\end{pmatrix} \ .
\end{aligned}
\end{equation}
The spectral flow mode is only modified by a global multiplication by the constant $z_{\alpha}^{1/2}$ meaning that $\ket{\psi_{-1}}$ is still an eigenstate of $\mathcal{H}$. In particular, the localization length (here equals to  $1$ in our units) is not affected by the different non-Hermitian terms considered.

%%%%%%%%%%%%%%%%%%%%%%%%%%%%%%%%%%%%%%%%%%%%%%%
\begin{figure*}[ht]
\centering
\includegraphics[width=1\textwidth]{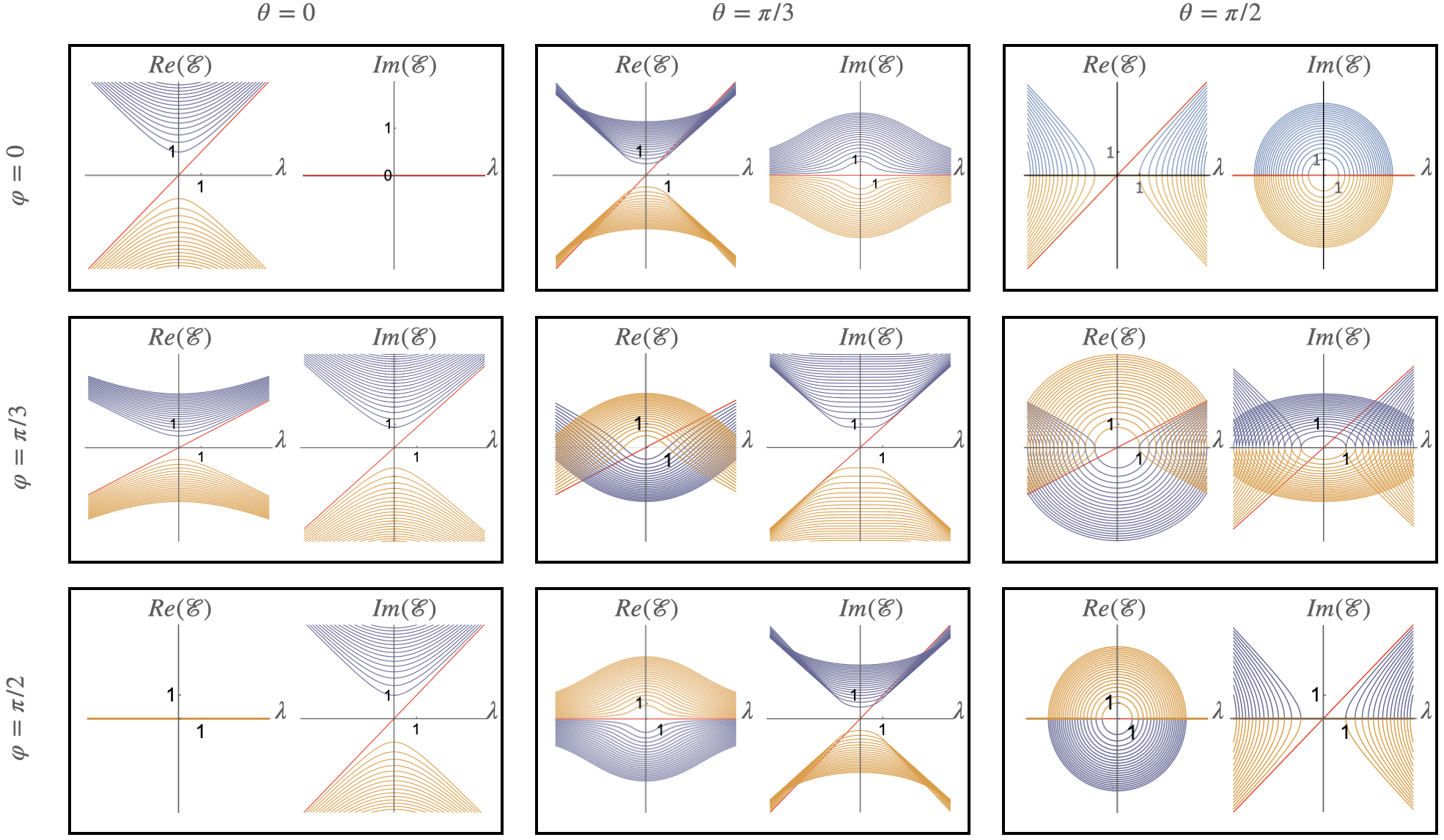}
\caption{\label{fig:full_spectrum} Real and imaginary parts of the spectrum $\Eop$ of $\mathcal{H}$, where $\tilde{\Eop}=\Eop/\sqrt{r_\alpha r_\beta}$ and $\tilde{\lambda}=\sqrt{r_\alpha r_\beta} \lambda$, and for $n$ up to $20$.
}
\end{figure*}

\section{Spectra of $\mathcal{H}$}
In this section we show in figure \ref{fig:full_spectrum} a few additional spectra of $\mathcal{H}$, for values of $\varphi$ different from zero . 

First, let us notice that, after the rescalings $\Eop\rightarrow\Eop/\sqrt{r_\alpha r_\beta}\equiv \tilde{\Eop}$ and $\lambda \rightarrow \sqrt{r_\alpha r_\beta} \lambda \equiv \tilde{\lambda}$, those spectra become independent of $r_\alpha$ and $r_\beta$.

As said in the main text, the spectral flow only appears in the real part of the spectrum when $\varphi=0$. Increasing $\varphi$ amounts to rotate the spectrum of an angle $\varphi$ in the complex plane. We see in figure \ref{fig:full_spectrum} that, as a consequence, both the real part and the imaginary part of the spectrum display a spectral flow, up to $\varphi=\pi/2$, where the spectral flow becomes purely imaginary. Indeed, the spectrum for $\theta=\pi/2$ is identical to that that for $\varphi=0$ up to a swapping of the real and imaginary parts.

%%%%%%%%%%%%%%%%%%%%%%%%%%%%%%%%%%%%%%%%%%%%%%%%
\section{Chern number of the symbol Hamiltonian}

\begin{figure}[ht]
    \centering
    \includegraphics[height=4cm]{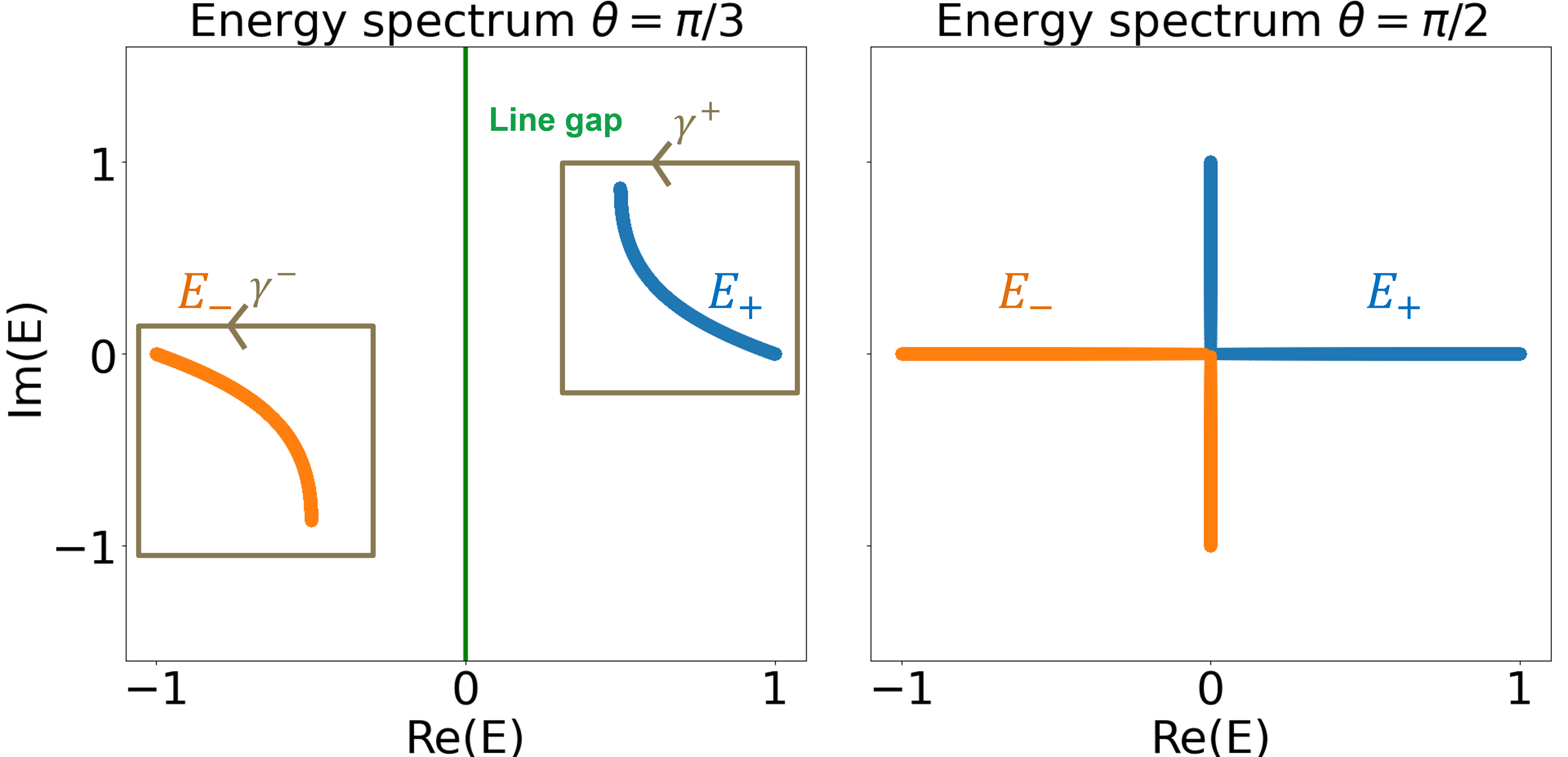}
    \caption{Spectrum of the symbol Hamiltonian $H_{\theta}[\lambda;x,p]$ for all $(\lambda,x,p) \in S^2$ and $\theta=\pi/3$ (left) where the positive and negative bands are separated by a line gap. For $\theta=\pi/2$ (right) at the contrary no line gap exist }
    \label{fig:line_gap}
\end{figure}

Let us consider the symbol Hamiltonian $H(\lambda;x,p)$ associated to the operator Hamiltonians $\mathcal{H}$. One can compute the spectrum of $H(\lambda;x,p)$ for each point on the sphere $(\lambda;x,p)\in S^2$. As $H(\lambda;x,p)$ acts on a vector space of finite spectrum, such spectrum has discrete bands which are parameterised by the choice of a point on  the sphere. The spectrum of $H(\lambda;x,p)$ is said to have a line gap if there are two bands in its spectrum which can be separated by a line in the complex plane (see Fig \ref{fig:line_gap}). 

Such a line gap exists for example for the symbol Hamiltonian
\begin{align}
H_{\theta}[\lambda;x,p] =  \frac{1}{\sqrt{2}} \begin{pmatrix}
\lambda  & e^{i\theta}(x- \ii p) \\
e^{i\theta}(x+\ii p)  &  -\lambda 
\end{pmatrix} 
\label{eq:Hprime}
\end{align}
when $e^{i2\theta}\neq -1$ (in other words $ \theta \neq \pm \pi/2$) as its spectrum 
\begin{equation}
    E_\pm = \pm \sqrt{\lambda^2+e^{i2\theta} (x^2+p^2)}
\end{equation}
is always separated by the vertical line $i\mathds{R} \subset \mathds{C}$ for all $(\lambda,x,p)$ in the sphere. In contrast, the case  $e^{i2\theta}=-1$ has no line gap as both energies are degenerated $E_+=E_-=0$ for $(\lambda=1,x=1,p=0)/\sqrt{2}$ (see Fig \ref{fig:line_gap}). 

When such a line gap exists, one can denote by $+/-$ the bands at the right/left of the separation line and then define the projectors $P^{\pm}(\lambda,x,p)$ on the eigenstates $\psi^\pm(x) = (1
,e^{-i\theta} \lambda-E_\pm/(x-ip))^t $ of such band $\pm$, with supplementary the eigenstates of the other band $\mp$. This can be done explicitly through the Cauchy formula
\begin{align}
  P^{\pm} = \frac{1}{2\ii \pi} \oint_{\gamma^{\pm}} \frac{\dd z}{H-z}  \label{eq:CauchyOperator}
\end{align}
where $\gamma^\pm$ is a path enclosing solely the band $E_\pm$ in the complex plane (which exists if and only if there is a line gap). 

If $H$ is a two by two matrix which can be diagonalised as
\begin{equation}
    H = MDM^{-1} = M \begin{pmatrix}
    E_+ &0 \\ 0& E_-
    \end{pmatrix} M^{-1}
\end{equation}
one can easily prove this fact by first checking that
\begin{equation}
    (H-z)^{-1} = M \begin{pmatrix}
    \frac{1}{E_+-z} &0 \\ 0&  \frac{1}{E_--z}
    \end{pmatrix} M^{-1} \label{eq:32}
\end{equation}
since
\begin{equation}
\begin{aligned}
 &(H-z)M \begin{pmatrix}
    \frac{1}{E_+-z} &0 \\ 0&  \frac{1}{E_--z}
    \end{pmatrix} M^{-1}\\
    =&M \begin{pmatrix}
   (E_+-z) &0 \\ 0&  (E_--z)
    \end{pmatrix} \begin{pmatrix}
    \frac{1}{E_+-z} &0 \\ 0&  \frac{1}{E_- -z}
    \end{pmatrix} M^{-1}\\
    =&\mathds{1}
\end{aligned}    
\end{equation}
which implies \eqref{eq:32} by unicity of the inverse.

Applying this property to the definition of the projectors \eqref{eq:CauchyOperator} for a path $\gamma^+$ enclosing only the eigenvalue $E_+$, one can apply the Cauchy formula for complex numbers and obtain 
\begin{equation}
\begin{aligned}
     P &= M \begin{pmatrix}
    \frac{1}{2i\pi} \oint_{\gamma^+} \frac{dz}{E_+-z}&0\\0&\frac{1}{2i\pi} \oint_{\gamma^+} \frac{dz}{E_--z}
    \end{pmatrix}M^{-1}\\
    &= M \begin{pmatrix}
   1&0\\0&0
    \end{pmatrix}M^{-1} \ .
\end{aligned}
\end{equation}
So, $P$ is an actual projector even if it is not an orthogonal one  $P^\dagger \neq P$. The only difference with an orthogonal projector is that it cannot be written in the form $P = \ket{\psi}\!\!\bra{\psi}$. Instead, if one denotes by $\ket{\psi}$ the right eigenvector of $H$ of eigenvalue $E_+$ and $\ket{\Tilde{\psi}}$ the left eigenvector of $H$ of eigenvalue $E_+$ (which is also  the right eigenvector of $H^\dagger$ of eigenvalue $E^*_+$, the complex conjugate of $E_+$) one has
\begin{equation}
    \ket{\psi}= M \begin{pmatrix}1\\0\end{pmatrix} \text{ and }\ket{\Tilde{\psi}}= (M^{-1})^\dagger \begin{pmatrix}1\\0\end{pmatrix}
\end{equation}
which yields
\begin{equation}
\begin{aligned}
P = M \begin{pmatrix}1\\0\end{pmatrix}\cdot\begin{pmatrix}1&0\end{pmatrix}M^{-1}=\ket{\psi}\!\!\bra{\Tilde{\psi}} \ .
\end{aligned}
\end{equation}
which is  the expression used in the main text.

The fact that we can construct the spectral projectors $P^\pm=\ket{\psi^{\pm}}\!\!\bra{\Tilde{\psi}^\pm}$ when the  Hamiltonian has well separated bands allows us  to extend the definition of the Chern-number to the non-Hermitian case as
\begin{align}
  \ch(P^{\pm})\equiv \frac{1}{2\ii \pi}\int_{S^2} \tr \,P^{\pm} \dd P^{\pm} \wedge \dd P^{\pm} \ .
\end{align}

\section{homotopic invariance of the Chern number}

In this section, we present a proof that the previously defined Chern number $\mathcal{C}_\theta$ is invariant under the deformation of any homotopic parameter $\theta$ which does not close the line gap. Therefore it implies that the Chern number remains a topological invariant in the non-Hermitian case. 

To do so, we assume that we have a family of Hamiltonians $\ham$ with associated symbols $H_\theta[\lambda;x,p]$ which have a line gap in their spectrum for $(\lambda,x,p) \in S^2$. As mentioned in the main text we can therefore construct a family of projectors $P_\theta^\pm[\lambda;x,p]$ for $(\lambda,x,p) \in S^2$. We can then define for each $\theta$ the following quantity 
\begin{align}
\mathcal{C}_{\theta}^\pm  \equiv \frac{1}{2\pi i} \int_{(\lambda,x,p) \in S^2} \tr(P_{\theta}^\pm \dd P_{\theta}^\pm \wedge \dd P_{\theta}^\pm)
\end{align}
that we demonstrate in what follows to be indeed a Chern number, by showing that
 $\partial_\theta \mathcal{C}_{\theta} = 0$. For that purpose, we first compute
\begin{equation}
\begin{aligned}
 \partial_\theta \mathcal{C}_{\theta} =&  \frac{1}{2\pi i} \int_{ S^2} \tr\left[(\partial_\theta P_{\theta}) \dd P_{\theta}  \dd P_{\theta}\right.\\
 &+ \left.P_{\theta} \dd (\partial_\theta P_{\theta})  \dd P_{\theta} +P_{\theta} \dd P_{\theta}  \dd  (\partial_\theta P_{\theta})\right]
\end{aligned}
\end{equation}

where $P_\theta$ is a short hand for $P_\theta^\pm$ and where we have set $\dd P \dd P \equiv \dd P \wedge \dd P$ for commodity.
Because the integration on closed surface of an exact form vanishes  by Stokes' theorem we can then perform an integration using that
\begin{equation}
\begin{aligned}
d(P_\theta \partial_\theta P_\theta dP_\theta) =& dP_\theta \partial_\theta P_\theta dP_\theta + P_\theta d(\partial_\theta P_\theta) dP_\theta\\
&+ \cancel{P_\theta \partial_\theta P_\theta d^2P_\theta}
\end{aligned}
\end{equation}
\begin{equation}
\begin{aligned}
d(P_\theta dP_\theta \partial_\theta P_\theta ) =& dP_\theta dP_\theta \partial_\theta P_\theta  + \cancel{P_\theta \partial_\theta d^2P_\theta P_\theta }\\
&- P_\theta  dP_\theta d(\partial_\theta P_\theta)
\end{aligned}
\end{equation}
where the crossed terms are cancelled because $d^2=0$ and the minus sign in the second expression comes from the specific Leibniz rule of the exterior derivative. These integrations by part then lead to
\begin{equation}
\begin{aligned}
\label{eq:decomp}
 \partial_\theta \mathcal{C}_{\theta} =&  \frac{1}{2\pi i} \int_{ S^2} \tr\left[(\partial_\theta P_{\theta}) \dd P_{\theta}  \dd P_{\theta}\right.\\
 &- \left. \dd P_{\theta} (\partial_\theta P_{\theta})  \dd P_{\theta} + \dd P_{\theta} \dd P_{\theta}  (\partial_\theta P_{\theta})\right] \ .
\end{aligned}
\end{equation}
In fact, all the terms in this sum have a vanishing trace. In order to show that, we can focus on the first term $\tr[(\partial_\theta P_{\theta}) \dd P_\theta \dd P_{\theta}]$ as the proof will be the same for the other. To do so, we  use two properties of a projector: First, the identity can be decomposed as $1 = P^2_\theta + (1-P_\theta)^2$ so that we can write
\begin{equation}
\begin{aligned}
&\tr[(\partial_\theta P_{\theta}) \dd P_{\theta}\dd P_{\theta}] =
\tr[(P^2_\theta+ (1-P_\theta)^2)  (\partial_\theta P_{\theta}) \dd P_{\theta}\dd P_{\theta}]\\
 =&\tr[P_\theta^2  (\partial_\theta P_{\theta}) \dd P_{\theta}\dd P_{\theta}+(1-P_\theta)^2(\partial_\theta P_{\theta}) P_\theta \dd P_{\theta}\dd P_{\theta}]
\end{aligned}
\end{equation}
Second, we use  $P^2_\theta =P_\theta \rightarrow P_\theta \partial P_\theta = \partial P_\theta (1-P_\theta)$ for any derivative $\partial$ (i-e: for $\partial_\theta$ or $\dd$) which gives us a commutation rule between $P_\theta$ and its derivative. Therefore, we  show that
\begin{equation}
\begin{aligned}
&\tr[(\partial_\theta P_{\theta}) \dd P_{\theta}\dd P_{\theta}]\\
=&\tr[P_\theta  (\partial_\theta P_{\theta})(1-P_\theta) \dd P_{\theta}\dd P_{\theta}+ (1-P_\theta)(\partial_\theta P_{\theta}) P_\theta \dd P_{\theta}\dd P_{\theta}]\\
 =&\tr[P_\theta  (\partial_\theta P_{\theta}) \dd P_{\theta}P_\theta\dd P_{\theta}+ (1-P_\theta)(\partial_\theta P_{\theta})  \dd P_{\theta}(1-P_\theta)\dd P_{\theta}]\\
 =&\tr[(1-P_\theta)P_\theta  (\partial_\theta P_{\theta}) \dd P_{\theta}\dd P_{\theta}+ (1-P_\theta)P_\theta(\partial_\theta P_{\theta})  \dd P_{\theta}\dd P_{\theta}]
\end{aligned}
\end{equation}
where the last line is equal to zero since $(1-P_\theta)P_\theta=0$.

We can  similarly show that all the terms in the expression \eqref{eq:decomp} of $\partial_\theta \mathcal{C}_{\theta}$ vanish. We therefore conclude that the Chern number is an homotopic invariant with $\theta$ as long as the spectral projectors $P_\theta$ are well defined, that is when there is a line gap.

\section{Invariance of the Chern number under the transformation (9)}

In the main text, the transformation \eqref{eq:transformA} helps us to reduce the number of relevant parameters  to a single one, i.e. $\theta$. Let us recall that this transformation can be written as
\begin{align}
    A^{-1} \mathcal{H}[\sqrt{r_\alpha r_\beta}\lambda] A = \sqrt{r_\alpha r_\beta}\ee^{\ii \varphi} \ham[\lambda] 
    \label{eq:transformAappendix2}
\end{align}
where $A$ is some invertible matrix independent of $(\lambda,x,p)$. We then compute the Chern number of $\mathcal{H}_\theta$ using \eqref{eq:ChernnonHerm} and link it to the number of topological edge modes. However, the definition of the Chern number \eqref{eq:ChernnonHerm} can also be applied to the initial Hamiltonian $\mathcal{H}$ and one can wonder if that would change anything. In fact in this section we quickly show that if  $\mathcal{H}$ and $\mathcal{H}_\theta$ are related by the transformation \eqref{eq:transformAappendix2}. So their Chern numbers are equal.

Let us analyse the expression of the transformation \eqref{eq:transformAappendix2}. First, the factor $\sqrt{r_\alpha r_\beta}\ee^{\ii \varphi}$ is just a re-scaling parameter in energy which does not alter the eigenstates of $H$ and therefore does not alter the definition of the band projectors $P^\pm=\ket{\psi^\pm}\!\!\bra{\psi^\pm}$. It thus does not modify the Chern numbers. 

If we analyse the rest of the transformation \eqref{eq:transformAappendix2}, one can see that it straightforwardly translates to the symbol Hamiltonians $H$ and $H_{\theta}$ (of $\mathcal{H}$ and $\mathcal{H}_\theta$), so that their projectors are related as 
\begin{align}
P^{\pm}[\lambda;x,p] = A P_\theta^{\pm}[\sqrt{r_\alpha r_\beta}\lambda;x,p]A^{-1}.
\end{align}

Since $A$ has no dependence in $(\lambda, x, p)$, then  the Berry curvature remains unchanged under the action of $A$ and we have
\begin{equation}
    \tr(P^{\pm}_{ \theta} \dd P^{\pm}_{ \theta}\wedge \dd P^{\pm}_{\theta}) = \tr(P^{\pm} \dd P^{\pm}\wedge \dd P^{\pm})
\end{equation}
up to a rescaling $\lambda \xrightarrow[]{}\sqrt{r_\alpha r_\beta}\lambda$.

Thus the expression of the Chern number $\mathcal{C}^{\pm}$ of $\mathcal{H}$ reduces to that of the Chern number $\mathcal{C}^{\pm}_\theta$ of $\mathcal{H}_\theta$, up to a deformation of the surface of integration $\mathcal{S}$ in the $\lambda$ direction that elongates the sphere into an ellipsoid. But as the Berry curvature is a closed form $\dd\left(P^{\pm}_{ \theta} \dd P^{\pm}_{ \theta}\wedge \dd P^{\pm}_{\theta}\right)=0$ (outside of the singularity $(\lambda,x,p)=0$), the Chern number is invariant under smooth changes of the integration surface. Therefore we deduce that the equality $\mathcal{C}^{\pm}=\mathcal{C}^{\pm}_\theta$ holds.

\section{A general spectral flow index and its homotopic invariance}

The goal of this section is to explain how the spectral flow index can be computed in general and to prove that it is a topological invariant as long as its symbol has a line gap for large $(\lambda,x,p)$. For this, we consider an operator Hamiltonian $\mathcal{H}[\lambda]$ whose symbol $H(\lambda,x,p)$ has a line gap for $(\lambda,x,p)\neq 0$. For simplification, we will also assume that the line gap is the vertical imaginary line, as any other direct gap in the complex plane is deduced from this situation by a rotation. 

\begin{figure}[ht]
    \centering
    \includegraphics[width = 7cm]{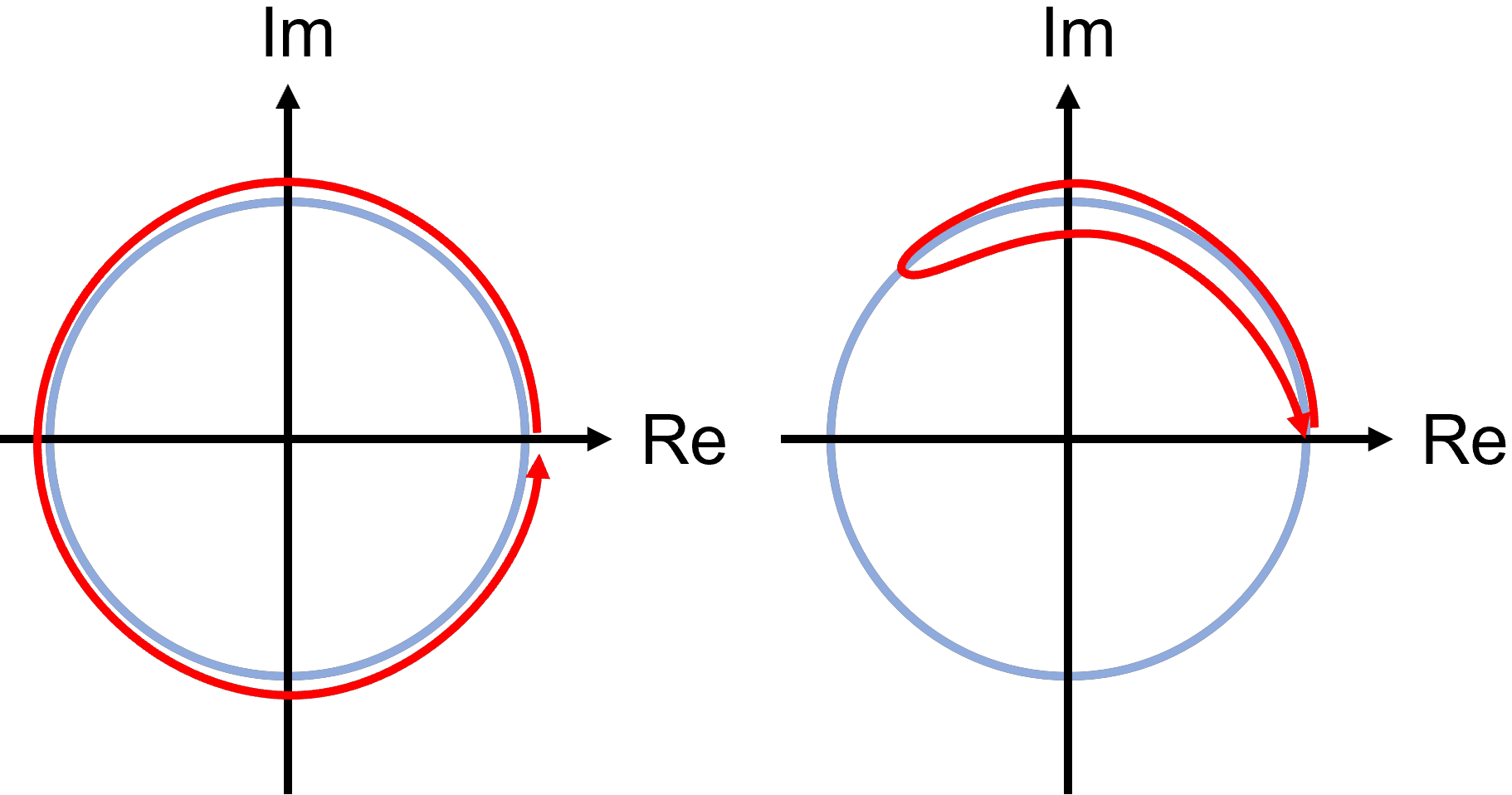}
    \caption{Sketches of paths followed by the complex eigenvalue $\exp(i2\pi/(1+e^{-\mathcal{E}_\lambda/\delta}))$ of $\mathcal{U}^{+}$ when $\lambda$ is varied from $-\infty$ to $+\infty$, for states of $\mathcal{H}[\lambda]$ whose energy $\mathcal{E}_\lambda$ crosses the gap from the negative to the positive band (left) or remains on the same side of the gap (right)}
    \label{fig:path}
\end{figure}

From the operator Hamiltonian, we define the operators 
\begin{equation}
\begin{aligned}
    \mathcal{P}^{\pm}[\lambda] &= \frac{1}{1+e^{\mp\mathcal{H}[\lambda]/\delta}} 
\end{aligned}
\end{equation}
where $\delta$ is a small arbitrary real number. 
When $\delta$ is much smaller that the amplitude of the real part of the line gap, the functions $1/(1+e^{\mp \mathcal{E}/\delta})$ forms plateaus of value $1$ and $0$ for energies  $-\text{Re}(\mathcal{E})\gg \delta$ or $\text{Re}(\mathcal{E})\gg \delta$.  Therefore, $\mathcal{P}^\pm[\lambda]$ coincide with the projectors of the gapped states with negative/positive $\text{Re}(\mathcal{E})$. In practice, the exact value of $\delta$ does not matter as long as it is much smaller than the amplitude of the real part of the line gap.
In contrast, the spectrum of $\mathcal{P}^\pm[\lambda]$ smoothly interpolates between $1$ and $0$ for the spectral flow modes which cross the line gap. This distinction can then be encoded in the \sout{unitary} operator 
\begin{align}
\mathcal{U}^\pm[\lambda]\equiv e^{i2\pi\mathcal{P}^\pm[\lambda]} 
\end{align}
which acts as the identity when $\lambda \rightarrow \pm \infty$. Therefore, when $\lambda$ is varied from $-\infty$ to $+\infty$, the eigenvalues $\exp(i2\pi/(1+e^{\mp \mathcal{E}_\lambda/\delta}))$ of $\mathcal{U}^\pm$ form close curves in the complex plane that go through $1$, as illustrated in figure \ref{fig:path}. Those loops wind around the origin if and only if they are associated to the eigenvalue $\mathcal{E}_\lambda$ of a spectral flow mode. Their winding number is $+1$ (resp. $-1$) for a gapless state that extends from the left (resp. right) side of the line gap for $\lambda \approx -\infty$  to the right (resp. left) side of the line gap for $\lambda \approx \infty$. In contrast, the loops associated with the gapped states have a winding number zero. In the case where $\mathcal{H}_0$ is Hermitian, those loops are confined on the unit circle.
Each gapless state that crosses the gap will thus contribute with a winding number, so that the total spectral flow $\mathcal{N}^\pm$ of $\mathcal{H}$ is given by the winding number of $\mathcal{U}^\pm[\lambda]$, that is
\begin{equation}
\begin{aligned}
    \mathcal{N}^\pm &= \frac{1}{2i\pi} \int d\lambda\, \tr\left(\mathcal{U}^\pm\right)^{ -1} \partial_\lambda \mathcal{U}^\pm\\
    &= \frac{1}{2i\pi} \int d\lambda \partial_\lambda \ln(\det(\mathcal{U}^\pm))
\end{aligned}
\end{equation}
For example when $\mathcal{H}[\lambda]$ has the particular form
\begin{align}
\hamo[\lambda] = \begin{pmatrix}
\lambda  & \cre \\
 \an  &  -\lambda 
\end{pmatrix}
=\an \sigma_- + \cre \sigma_+ + \lambda \sigma_z
\end{align}
one can choose $\delta\ll 1$ so that the only gapless modes of $\hamo$ are the one of the form $\ket{\psi} \in \ker(\mathcal{D}=\an \sigma_-)$ and $\ket{\psi'} \in \ker(\mathcal{D}^\dagger)$. So, the only variation of $\mathcal{U}$ will occur in this basis which simplifies the computation and it can therefore be checked by hand that in this case 
\begin{equation}
    \mathcal{N}^{\pm}  = \pm(\text{dim Ker }\mathcal{D}- \text{dim Ker }\mathcal{D}^\dagger) = \pm \text{ind }\mathcal{D}.
\end{equation}

An important property of the spectral flow is that, as for the Chern number, it is a topological invariant, which means that it is left unchanged against deformations that preserve the line-gap of the symbol. For show that, consider a family of Hamiltonians $H_\theta[\lambda]$ paramatrised by a value $\theta$ and let us define their associated spectral flow $\mathcal{N}^{\pm}$. Then, one has
\begin{equation}
\begin{aligned}
    \partial_\theta \mathcal{N}^{\pm}_\theta &= \frac{1}{2i\pi} \int d\lambda \tr(\partial_\theta(\mathcal{U}^{\pm})^{-1} \partial_\lambda \mathcal{U}^\pm+ (\mathcal{U}^{\pm})^{-1} \partial_\theta\partial_\lambda \mathcal{U}^\pm)\\
    &= \frac{1}{2i\pi} \int d\lambda \tr(\partial_\theta(\mathcal{U}^{\pm})^{-1} \partial_\lambda \mathcal{U}^\pm-\partial_\lambda(\mathcal{U}^{\pm})^{-1} \partial_\theta \mathcal{U}^\pm)
\end{aligned}
\end{equation}
We then use that $\partial_\theta(\mathcal{U}^{\pm})^{-1}=- (\mathcal{U}^{\pm})^{-1} \partial_\theta\mathcal{U}^{\pm} (\mathcal{U}^{\pm})^{-1}$ to obtain
\begin{equation}
\begin{aligned}
    \partial_\theta \mathcal{N}^{\pm}_\theta =& -\frac{1}{2i\pi} \int d\lambda \tr((\mathcal{U}^{\pm})^{-1}\partial_\theta\mathcal{U}^{\pm}(\mathcal{U}^{\pm})^{-1} \partial_\lambda \mathcal{U}^\pm\\
    &- (\mathcal{U}^{\pm})^{-1}\partial_\lambda\mathcal{U}^{\pm}(\mathcal{U}^{\pm})^{-1} \partial_\theta \mathcal{U}^\pm)
\end{aligned}
\end{equation}
which is equal to zero due to the cyclicity of the trace. This shows that for two Hamiltonians in the same homotopy class, their spectral flow must be the same. Thus the spectral flow is a topological invariant. 

One should note that even if hidden in this demonstration, the assumption that the symbol Hamiltonian has a line gap for large $(\lambda,x,p)$ insures that $\mathcal{H}$ is gapped far from the interface and for large $\lambda$, which implies that $\partial_\theta \mathcal{U}^\pm \approx0$ in those far away regions. This property is crucial in order to have an integration by part in $\lambda$ without boundary terms in $\lambda = \pm \infty$ as well as being able to use the cyclicity of the trace.

\section{Pseudo-Hermitian to Hermitian mapping and $\lambda \leftrightarrow E$ swapping }

In this section we present a mapping  between a pseudo-Hermitian Hamiltonian and a Hermitian one that allows us to interpret a symmetry breaking phase of the first one as a spectral gap of the second one.

Consider a parameterised Hamiltonian $H(\lambda)$ with a pseudo-Hermitian symmetry, i.e. there exists a unitary operator $\sigma^\dagger=\sigma^{-1}$ such that
\begin{equation}
    \sigma^\dagger H(\lambda) \sigma  = H^\dagger(\lambda)\ .
    \label{eq:pseudoHermitian}
\end{equation}
This symmetry implies that if $\ket{\psi}$ is an eigenstate of $H(\lambda)$ with eigenvalue $E$, then $\sigma^\dagger\ket{\psi} \equiv \ket{\psi'}$ is a left eigenstate of $H(\lambda)$  with eigenvalue $E^*$ as 
\begin{equation}
    \bra{\psi}\sigma H = (H^\dagger\sigma^\dagger \ket{\psi})^\dagger= (\sigma^\dagger H\ket{\psi})^\dagger =\bra{\psi}\sigma E^* \, .
\end{equation}
Therefore as the left and right spectrum are equal, it implies that the eigenvalues in the spectrum of $H$ must either be real or appears in conjugates pairs $(E,E^*)$. When an eigenvalue is real, we say it "preserves" the pseudo-Hermitian symmetry whereas when the eigenvalues appear in conjugate pairs, we say they "spontaneously break" the pseudo-Hermitian symmetry \cite{NonHermiPhysics, SymProtectedDelplace}. When $H$ is parameterised by some parameter $\lambda$, the eigenvalues can generally reach a transition point where they switch from the symmetry preserving case, where they are real, to the spontaneously broken one where they appear in conjugated pairs (or vice versa).

Now that we have introduce the basic property of the pseudo-hermitian symmetry, let us assume that the parameter $\lambda$ is  ''conjugated'' to the symmetry operator $\sigma$, meaning that $H(\lambda)$ can be written as
\begin{equation}
    H(\lambda) = H + \lambda \sigma 
\end{equation}
(where $H$ is short for $H(\lambda=0)$). Then, the eigenvalue equation  
\begin{equation}
    ( H + \lambda \sigma)\ket{\psi}=E \ket{\psi}
    \label{eq:eigenE}
\end{equation}
comes together with another eigenvalue equation for $\ket{\psi'}=-\sigma^\dagger\ket{\psi}$
\begin{equation}
    ( -H\sigma + E \sigma)\ket{\psi'}=\lambda \ket{\psi'} \ .
    \label{eq:eigenL}
\end{equation}
Importantly, the roles of the parameter and of the eigenvalue have been swapped between Eq \eqref{eq:eigenE} and Eq \eqref{eq:eigenL}.

Writing $\Tilde{H}(E)\equiv -H\sigma +E\sigma$, and using the pseudo-Hermitian symmetry \eqref{eq:pseudoHermitian} for $H$, we can write
\begin{align}
    \Tilde{H}^\dagger(E) = -H\sigma^\dagger +E^*\sigma^\dagger
\end{align}
Therefore, $\Tilde{H}(E)$ is Hermitian provided that $E$ is real and that the unitary operator of the pseudo-Hermitian symmetry is also Hermitian, that is $\sigma=\sigma^\dagger$. 
In other words, the initial eigenenergy problem in $E$ for the pseudo-Hermitian Hamiltonian $H(\lambda)$ with $\sigma=\sigma^{-1}=\sigma^\dagger$, is in one-to-one correspondence with the eigenvalue problem in $\lambda$ for $\Tilde{H}(E)$. Considering $E$ real thus implies the Hermiticity of $\Tilde{H}(E)$ and thus guarantees the realness of $\lambda$.

When we impose the constrain that $E$ must be real, we see that in the $\Tilde{H}$ eigenvalue problem we only keep  the modes of $H$ which have a real energy and dismiss those of the spontaneously broken phase. Therefore the transition toward such a symmetry broken phase in $\lambda$ is now reinterpreted as a transition toward a region of the spectrum of $\Tilde{H}(E)$ which is gapped in the Hermitian picture. So the recast of a pseudo-Hermitian eigenvalue problem into an Hermitian one is quite powerful and shed new light in the spontaneously broken phase transition by reinterpreting it in a gap transition and vice versa.


\begin{thebibliography}{10}

\bibitem{TopoBandNonHerm}
Huitao Shen, Bo~Zhen, and Liang Fu.
\newblock Topological band theory for non-hermitian hamiltonians.
\newblock {\em Phys. Rev. Lett.}, 120:146402, Apr 2018.

\bibitem{NewNonHermiInvariants}
Ananya Ghatak and Tanmoy Das.
\newblock New topological invariants in non-hermitian systems.
\newblock {\em Journal of Physics: Condensed Matter}, 31(26):263001, Apr 2019.

\bibitem{TopoPhase}
Zongping Gong, Yuto Ashida, Kohei Kawabata, Kazuaki Takasan, Sho Higashikawa,
  and Masahito Ueda.
\newblock Topological phases of non-hermitian systems.
\newblock {\em Phys. Rev. X}, 8:031079, Sep 2018.

\bibitem{lutopological2014}
Ling Lu, John~D. Joannopoulos, and Marin Soljačić.
\newblock Topological photonics.
\newblock {\em Nature Photonics}, 8(11):821--829, November 2014.

\bibitem{Nash14495}
Lisa~M. Nash, Dustin Kleckner, Alismari Read, Vincenzo Vitelli, Ari~M. Turner,
  and William T.~M. Irvine.
\newblock Topological mechanics of gyroscopic metamaterials.
\newblock {\em Proceedings of the National Academy of Sciences},
  112(47):14495--14500, 2015.

\bibitem{Delplace_2017}
Pierre Delplace, J.~B. Marston, and Antoine Venaille.
\newblock Topological origin of equatorial waves.
\newblock {\em Science}, 358(6366):1075--1077, oct 2017.

\bibitem{Shankar_2022}
Suraj Shankar, Anton Souslov, Mark~J. Bowick, M.~Cristina Marchetti, and
  Vincenzo Vitelli.
\newblock Topological active matter.
\newblock {\em Nature Reviews Physics}, 4(6):380--398, may 2022.

\bibitem{Zhang_2018}
Xiujuan Zhang, Meng Xiao, Ying Cheng, Ming-Hui Lu, and Johan Christensen.
\newblock Topological sound.
\newblock {\em Communications Physics}, 1(1), dec 2018.

\bibitem{parker_2021}
Jeffrey~B. Parker.
\newblock Topological phase in plasma physics.
\newblock {\em Journal of Plasma Physics}, 87(2):835870202, 2021.

\bibitem{weimann_topologically_2017}
S.~Weimann, M.~Kremer, Y.~Plotnik, Y.~Lumer, S.~Nolte, K.~G. Makris, M.~Segev,
  M. C. Rechtsman, and A.~Szameit.
\newblock Topologically protected bound states in photonic
  parity–time-symmetric crystals.
\newblock {\em Nature Materials}, 16(4):433--438, April 2017.

\bibitem{GainLossExperiment}
Shuo Liu, Shaojie Ma, Cheng Yang, Lei Zhang, Wenlong Gao, Yuan~Jiang Xiang,
  Tie~Jun Cui, and Shuang Zhang.
\newblock Gain- and loss-induced topological insulating phase in a
  non-hermitian electrical circuit.
\newblock {\em Physical Review Applied}, 13(1), Jan 2020.

\bibitem{NonHermiPhysics}
Yuto Ashida, Zongping Gong, and Masahito Ueda.
\newblock Non-hermitian physics.
\newblock {\em Advances in Physics}, 69(3):249–435, Jul 2020.

\bibitem{TopologicalLaser}
Miguel~A. Bandres, Steffen Wittek, Gal Harari, Midya Parto, Jinhan Ren,
  Mordechai Segev, Demetrios~N. Christodoulides, and Mercedeh Khajavikhan.
\newblock Topological insulator laser: Experiments.
\newblock {\em Science}, 359(6381), 2018.

\bibitem{NonlinearReview}
Daria Smirnova, Daniel Leykam, Yidong Chong, and Yuri Kivshar.
\newblock Nonlinear topological photonics.
\newblock {\em Applied Physics Reviews}, 7(2):021306, 2020.

\bibitem{Longhi2016}
Stefano Longhi.
\newblock Non-hermitian tight-binding network engineering.
\newblock {\em Physical Review A}, 93(2), feb 2016.

\bibitem{TonyLee16}
Tony~E. Lee.
\newblock Anomalous edge state in a non-hermitian lattice.
\newblock {\em Phys. Rev. Lett.}, 116:133903, Apr 2016.

\bibitem{xiong17}
Ye~Xiong.
\newblock Why does bulk boundary correspondence fail in some non-hermitian
  topological models.
\newblock {\em arXiv:1705.06039}, 2017.

\bibitem{Yao2018}
Shunyu Yao and Zhong Wang.
\newblock Edge states and topological invariants of non-hermitian systems.
\newblock {\em Phys. Rev. Lett.}, 121:086803, Aug 2018.

\bibitem{NonHemChernBands}
Shunyu Yao, Fei Song, and Zhong Wang.
\newblock Non-hermitian chern bands.
\newblock {\em Phys. Rev. Lett.}, 121:136802, Sep 2018.

\bibitem{Brzezicki19}
Wojciech Brzezicki and Timo Hyart.
\newblock Hidden chern number in one-dimensional non-hermitian chiral-symmetric
  systems.
\newblock {\em Phys. Rev. B}, 100:161105, Oct 2019.

\bibitem{Deng19}
Tian-Shu Deng and Wei Yi.
\newblock Non-bloch topological invariants in a non-hermitian domain wall
  system.
\newblock {\em Phys. Rev. B}, 100:035102, Jul 2019.

\bibitem{Borgnia20}
Dan~S. Borgnia, Alex~Jura Kruchkov, and Robert-Jan Slager.
\newblock Non-hermitian boundary modes and topology.
\newblock {\em Phys. Rev. Lett.}, 124:056802, Feb 2020.

\bibitem{guo2021analysis}
Gang-Feng Guo, Xi-Xi Bao, and Lei Tan.
\newblock The analysis of bulk boundary correspondence under the singularity of
  the generalized brillouin zone in non-hermitian system.
\newblock {\em arXiv preprint arXiv:2106.06384}, 2021.

\bibitem{Yang2020}
Zhesen Yang, Kai Zhang, Chen Fang, and Jiangping Hu.
\newblock Non-hermitian bulk-boundary correspondence and auxiliary generalized
  brillouin zone theory.
\newblock {\em Phys. Rev. Lett.}, 125:226402, Nov 2020.

\bibitem{Kunst18}
Flore~K. Kunst, Elisabet Edvardsson, Jan~Carl Budich, and Emil~J. Bergholtz.
\newblock Biorthogonal bulk-boundary correspondence in non-hermitian systems.
\newblock {\em Phys. Rev. Lett.}, 121:026808, Jul 2018.

\bibitem{Edvardsson19}
Elisabet Edvardsson, Flore~K. Kunst, and Emil~J. Bergholtz.
\newblock Non-hermitian extensions of higher-order topological phases and their
  biorthogonal bulk-boundary correspondence.
\newblock {\em Phys. Rev. B}, 99:081302, Feb 2019.

\bibitem{RestoBulkBoundary}
Matteo Brunelli, Clara~C. Wanjura, and Andreas Nunnenkamp.
\newblock Restoration of the non-hermitian bulk-boundary correspondence via
  topological amplification, 2022.

\bibitem{Song_2019}
Fei Song, Shunyu Yao, and Zhong Wang.
\newblock Non-hermitian topological invariants in real space.
\newblock {\em Phys. Rev. Lett.}, 123:246801, Dec 2019.

\bibitem{Type34nonhermi}
Zaur~Z. Alisultanov and Edvin~G. Idrisov.
\newblock Towards the theory of types iii and iv non-hermitian weyl fermions,
  2021.

\bibitem{VolovikBook}
G.~E. Volovik.
\newblock {\em The Universe in a Helium Droplet}.
\newblock OUP Oxford, 2009.

\bibitem{BerryChernMonopoles}
Pierre Delplace.
\newblock {Berry-Chern monopoles and spectral flows}.
\newblock {\em SciPost Phys. Lect. Notes}, page~39, 2022.

\bibitem{Faure2019}
Frédéric Faure.
\newblock Manifestation of the topological index formula in quantum waves and
  geophysical waves, 2019.

\bibitem{Venaille2022}
Antoine Venaille, Yohei Onuki, Nicolas Perez, and Armand Leclerc.
\newblock From ray tracing to topological waves in continuous media, 2022.

\bibitem{Bellissard95}
Jean Bellisard.
\newblock Change of the chern number at band crossings.
\newblock {\em arXiv:cond-mat/9504030v1}, 1995.

\bibitem{touchais22}
Jean-Baptiste Touchais, Pascal Simon, and Andrej Mesaros.
\newblock Robust propagating in-gap modes due to spin-orbit domain walls in
  graphene.
\newblock {\em Phys. Rev. B}, 106:035139, Jul 2022.

\bibitem{upreti20}
Lavi~K. Upreti and Pierre Delplace.
\newblock Topological chiral interface states beyond insulators.
\newblock {\em Phys. Rev. A}, 102:023520, Aug 2020.

\bibitem{perrot2019topological}
Manolis Perrot, Pierre Delplace, and Antoine Venaille.
\newblock Topological transition in stratified fluids.
\newblock {\em Nature Physics}, 15(8):781--784, 2019.

\bibitem{Marciani_2020}
M.~Marciani and P.~Delplace.
\newblock Chiral maxwell waves in continuous media from berry monopoles.
\newblock {\em Physical Review A}, 101(2), feb 2020.

\bibitem{venaille21}
A.~Venaille and P.~Delplace.
\newblock Wave topology brought to the coast.
\newblock {\em Phys. Rev. Research}, 3:043002, Oct 2021.

\bibitem{zhu2021topology}
Ziyan Zhu, Christopher Li, and J.~B. Marston.
\newblock Topology of rotating stratified fluids with and without background
  shear flow, 2021.

\bibitem{Langmuircyclotron}
Hong Qin and Yichen Fu.
\newblock Topological langmuir-cyclotron wave, 2022.

\bibitem{PerezPRL2022}
Nicolas Perez, Pierre Delplace, and Antoine Venaille.
\newblock Unidirectional modes induced by nontraditional coriolis force in
  stratified fluids.
\newblock {\em Phys. Rev. Lett.}, 128:184501, May 2022.

\bibitem{iwai_topological_2014}
T.~Iwai and B.~Zhilinskii.
\newblock Topological phase transitions in the vibration–rotation dynamics of
  an isolated molecule.
\newblock {\em Theoretical Chemistry Accounts}, 133(7):1501, May 2014.

\bibitem{iwai_2016}
T.~Iwai and B.~Zhilinskii.
\newblock Band rearrangement through the 2d-dirac equation: Comparing the aps
  and the chiral bag boundary conditions.
\newblock {\em Indagationes Mathematicae}, 27(5):1081--1106, 2016.
\newblock Dynamics and Geometry.

\bibitem{FZ2000}
F.~Faure and B.~Zhilinskii.
\newblock Topological chern indices in molecular spectra.
\newblock {\em Phys. Rev. Lett.}, 85:960--963, Jul 2000.

\bibitem{FZ2001}
F.~Faure and B.~Zhilinskii.
\newblock Topological properties of the born–oppenheimer approximation and
  implications for the exact spectrum.
\newblock {\em Letters in Mathematical Physics}, 55:219, 2001.

\bibitem{Zyuzin_2012}
A.~A. Zyuzin and A.~A. Burkov.
\newblock Topological response in weyl semimetals and the chiral anomaly.
\newblock {\em Physical Review B}, 86(11), sep 2012.

\bibitem{Burkov_2015}
A~A Burkov.
\newblock Chiral anomaly and transport in weyl metals.
\newblock {\em Journal of Physics: Condensed Matter}, 27(11):113201, feb 2015.

\bibitem{Bradlyn2016}
Barry Bradlyn, Jennifer Cano, Zhijun Wang, M.~G. Vergniory, C.~Felser, R.~J.
  Cava, and B.~Andrei Bernevig.
\newblock Beyond dirac and weyl fermions: Unconventional quasiparticles in
  conventional crystals.
\newblock {\em Science}, 353(6299), 2016.

\bibitem{Ezawa2017}
Motohiko Ezawa.
\newblock Chiral anomaly enhancement and photoirradiation effects in multiband
  touching fermion systems.
\newblock {\em Phys. Rev. B}, 95:205201, May 2017.

\bibitem{Avron1989}
J.~E. Avron, L.~Sadun, J.~Segert, and B.~Simon.
\newblock Chern numbers, quaternions, and berry's phases in fermi systems.
\newblock {\em Commun. Math. Phys.}, 124:595--627, 1989.

\bibitem{wan2011}
Xiangang Wan, Ari~M. Turner, Ashvin Vishwanath, and Sergey~Y. Savrasov.
\newblock Topological semimetal and fermi-arc surface states in the electronic
  structure of pyrochlore iridates.
\newblock {\em Phys. Rev. B}, 83:205101, May 2011.

\bibitem{atiyah1968index}
Michael~Francis Atiyah and Isadore~Manuel Singer.
\newblock The index of elliptic operators: I.
\newblock {\em Annals of mathematics}, pages 484--530, 1968.

\bibitem{nakahara2018geometry}
Mikio Nakahara.
\newblock {\em Geometry, topology and physics}.
\newblock CRC press, 2018.

\bibitem{RussellYang22}
Russell Yang, Jun~Wei Tan, Tommy Tai, Jin~Ming Koh, Linhu Li, Stefano Longhi,
  and Ching~Hua Lee.
\newblock Designing non-hermitian real spectra through electrostatics.
\newblock {\em arXiv:2201.04153}, 2022.

\bibitem{Realspecwithnosym}
Kohei Kawabata and Masatoshi Sato.
\newblock Real spectra in non-hermitian topological insulators.
\newblock {\em Phys. Rev. Research}, 2:033391, Sep 2020.

\bibitem{WeylChern}
Yong Xu, Sheng-Tao Wang, and L.-M. Duan.
\newblock Weyl exceptional rings in a three-dimensional dissipative cold atomic
  gas.
\newblock {\em Phys. Rev. Lett.}, 118:045701, Jan 2017.

\bibitem{SymProtectedDelplace}
Pierre Delplace, Tsuneya Yoshida, and Yasuhiro Hatsugai.
\newblock Symmetry-protected multifold exceptional points and their topological
  characterization.
\newblock {\em Phys. Rev. Lett.}, 127:186602, Oct 2021.

\end{thebibliography}
\end{document}